# Intonation Perception in Real and Synthetic Speech across Varying Familiarity Levels: A Pilot Study of Equivalence Assessment

*Hanrui Zhou*[1,*], *Gaoyuan Zhang*[1,*], *Yixiang Chen*[1], *Yujie Xing*[2], *Feng Xu*[1], *Xurong Xie*[1,**], *Hui Chen*[1,**]

[1] Institute of Software, Chinese Academy of Sciences, Beijing, China
[2] Capital Normal University

annxx_zhr@163.com, zhanggaoyuan2023@iscas.ac.cn, chenyixiang24@mails.ucas.ac.cn, xingyujie010211@163.com, feng.xu2@hdr.mq.edu.au, xurong@iscas.ac.cn, chenhui@iscas.ac.cn

## Abstract

Language training relies on a corpus constructed by a large number linguistic materials. AI-powered voice clones provide a way to construct the corpus with relatively low cost. Singing voice conversion (SVC) model is used to generate synthetic voices. This study compares participants' performances on natural and synthetic speech in two experiments, similarity perception and intonation recognition. In the accuracy of similarity perception task, a significant interaction between speech type and intonation is found, suggesting that question may serve as a cue for speaker identification but may be influenced by synthetic features. In the accuracy of intonation recognition task, a significant interaction between speech type and familiarity is observed, indicating that speech type affects how much familiarity contributes to voice processing.

**Index Terms**: AI-powered voice clones, intonation perception, speaker familiarity

## 1. Introduction

In recent years, there has been growing recognition of the vital links between verbal communication and people's social and individual functioning. Language training serves as a fundamental approach on improving communicative ability, particularly in areas such as language acquisition and intervention for language disorders. One of the important factors for language training is the construction of high-quality corpora that contain specific linguistic data. However, such corpora are currently scarce [1]. An ideal high-quality corpus should encompass rich linguistic information, including speakers' identity and prosodic features (e.g., intonation, emotion prosody), to support the complex cognitive processes involved in language learning [2]. Yet building such a corpus requires substantial investment making it both economically and temporally costly, such as demanding skilled actors, controlled recording environments, and large data volumes. Thus, the lack of well-resourced corpora poses a significant challenge to advancing both research and practice in language training.

With the rapid development of AI voice cloning technology, it has become increasingly feasible to generate high-quality linguistic materials for speech training using synthetic voices. Recent studies have primarily focused on evaluating synthetic speech in terms of intelligibility, naturalness and speaker identity similarity [3, 4]. It has been shown that people are poorly equipped to detect AI-powered voice clones [5]. Research has shown that people are often unable to reliably distinguish AI-generated voices from human speech [5], and although listeners perform slightly better at voice discrimination than at image discrimination, they are still deceived most of the time [6, 7, 8]. However, most existing evaluations are based on declarative sentences [3, 4, 5, 6, 7, 8]. In Mandarin, question intonation is sometimes unmarked lexically. Intonation plays a central role in signaling question, particularly through a finally rising or high-level $f_0$ contour [9, 10, 11]. Intonation is a key component of natural speech, contributing significantly to its acoustic complexity [9], and remains a persistent challenge for voice cloning systems to fully replicate [12, 13]. Given that it is frequent in daily communication and constitute an essential part of language training, it is crucial to assess whether current AI-generated voices can accurately produce such prosodic distinctions.

There is evidence shown that familiarity with a talker's characteristics help facilitate linguistic processing [14]. Familiar voices help people identify the speakers and promote recovery of acoustic properties. Some studies attribute it to adaption. Speech perception is of adaption, difference in response between learned knowledge and strange stimulus or new acoustic patterns that haven't be learned before [15]. Adaptation effects often manifest as reduced reaction times and more stable categorization following brief exposure. In intervention contexts, repeated exposure is inherent to training paradigms [2]. Therefore, examining familiarity effects is crucial for evaluating the practical feasibility of AI-powered voice clones.

To address the gap, the present study compares 2 speech types (synthetic and natural voices) on different intonation (statement and question) and familiarity (familiar and strange) levels in two experiments, speaker identity similarity perception and intonation recognition. Specifically, this study sought to answer the following research questions: (1) How and to what extent does familiarity, intonation and speech type affect participants' performance of voice processing in speaker identity similarity perception task? (2) How do familiarity, intonation and speech type modulate participants' performance during intonation recognition.

The study makes two focused contributions: practically, preliminarily supporting the feasibility of incorporating AI-generated speech in corpus; theoretically, exploring cognitive differences in familiar/unfamiliar voice processing and in natural/AI-generated voice processing.

## 2. Methodology

### 2.1. Participant

A total of 17 participants are recruited. All of them are involved in speaker identity similarity perception, and 11 of them take part in intonation recognition. Listeners ages range from 22 - 41 years old (mean = 28.176, SD = 5.177), with 6 females and 11 males. The participants are native speakers of Mandarin

[*] These authors contributed equally.

[**] Corresponding authors.

Chinese, without hearing disorders. All participants have signed informed consent prior to participation in accordance with the guidelines of Institution of Software Chinese Academy of Sciences.

The participants are given an overview of the experiment then listen to natural voices conducted by two of their familiar fellows. Participants who fail to recognize the speaker will be removed from the analysis of speaker identity similarity perception.

## 2.2. Stimuli

### *2.2.1. Material*

A set of 11 neutral Mandarin sentences with 6 words are used. One of them is used for practice session, the other 10 are used in formal experiments. Examples include "爸爸打开大门" (My father is opening the door, /pa$^{51}$ pa$^{51}$ ta$^{214}$ kʰai$^{55}$ ta$^{51}$ mən$^{35}$/). 2 strangers and 2 people familiar with the participants are invited to read these sentences. This study adopts a binary operationalization of familiarity: "Strange" (voices from strangers) vs. "Familiar" (voices from acquaintances). All of the sentences are recorded twice in both statement and question intonations. Recordings with better quality are selected as the materials. Thus, the stimuli would not introduce unnaturalness that could be used as a cue for recognition. Meanwhile, these speakers are asked to recorded a speech with maximized emotional expressiveness for conversion of subsequent synthetic voices, without presenting to participants [16]. All speakers are native Chinese speakers, female, aging from 25 to 45 year. The recordings take place in a soundproof recording booth at a phonetic lab with Praat at 44.1 kHz. Figure 1 shows representative $f_0$ contours for "爸爸打开大门". The recording equipment used is RODE Wireless GO II microphone.

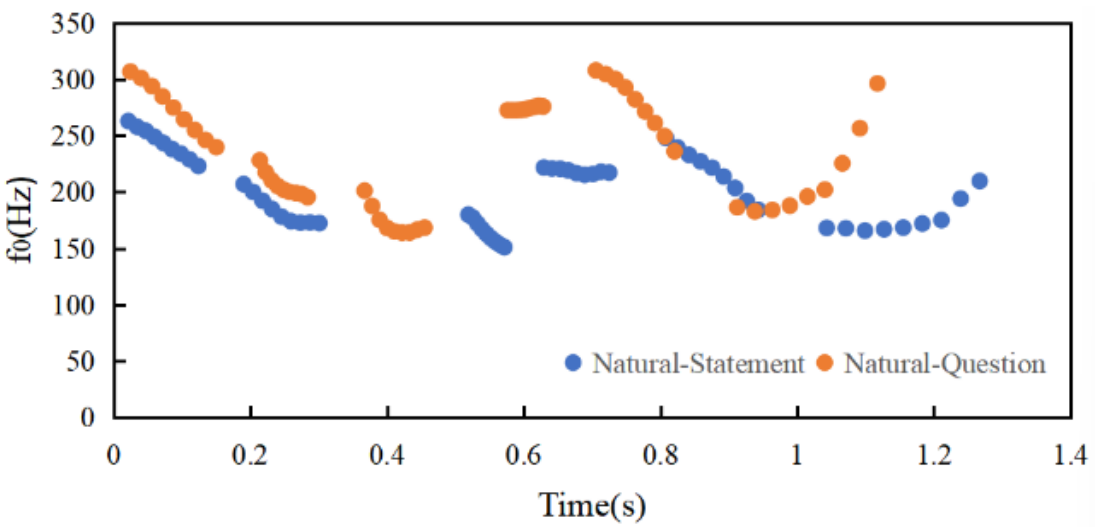

Figure 1: *Representative $f_0$ contours of "爸爸打开大门" produced as a natural statement and as a natural question*

### *2.2.2. AI-generated stimuli*

Synthetic voices are generated with an F0-conditioned singing voice conversion (SVC) model[1]. The core mechanism of the original SVC model involves capturing the semantic content and pitch variations (i.e., the F0 contour) from the source speaker's singing voice and converting them into the target speaker's timbre [17, 18].

To objectively evaluate the similarity between the converted voice and the target speaker's reference voice, we employ a widely-used, open-source speaker recognition model. Specifically, we utilize its speaker encoder to extract speaker embeddings from the audio segments. The cosine similarity between the embedding of the converted speech and that of the target speaker's reference speech is then computed and used as the speaker similarity metric. This score ranges from -1 to 1, with a higher value indicating greater similarity [19, 20]. Statement showed a mean similarity score of 0.55 (SD = 0.28), while questions showed a mean of 0.51 (SD = 0.30), suggesting comparable similarity levels across sentence types.

Synthetic voices of each sentence in higher score are selected. Then, 2 experts in phonetics are invited to pick the voices with the best intonation expression. If disagreement exists, the third expert is invited to evaluate and make selection. Figure 2 shows representative $f_0$ contours of AI-generated for "爸爸打开大门".

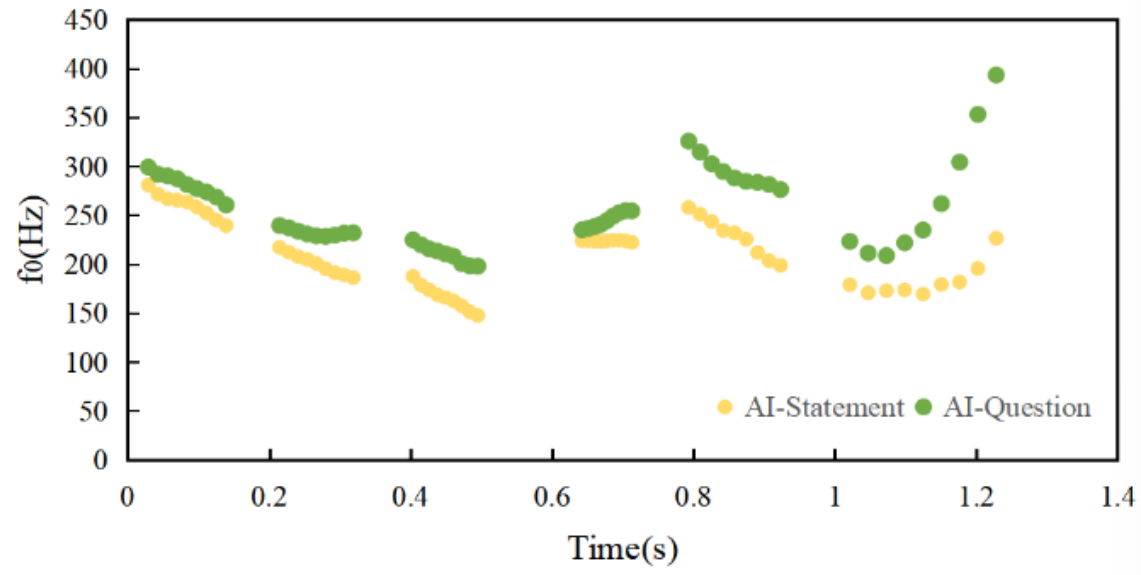

Figure 2: *Representative $f_0$ contours of "爸爸打开大门" produced as an AI statement and as an AI question*

## 2.3. Procedure

Generally, the study employs a 2 × 2 × 2 within-subject design, with the factor of speech type (natural vs. synthetic), familiarity to the speaker (familiar vs. strange) and intonation (statement vs. question). The dependent variables are response time (RT) and accuracy (ACC). The experiment is implemented in PsychoPy. The stimulus lists are all counterbalanced. The study is consisted of 2 experiments, speaker identity similarity perception and intonation recognition, conducting on different days. Before formal experiment, participants need to go through a practice session to ensure that they have understood the task. Sentences used in practice session do not present in the formal experiment.

### *2.3.1. Speaker identity similarity perception*

For speaker identity similarity perception, participants are asked to make an assessment on two different sentences they heard, choosing whether the sentences are conducted by the same person, then giving a score from 1 to 7 for the similarity. Point 7 accounts for that the two sentences in one trial are conducted by completely the same person, point 1 indicated completely different persons. Inspired by the experimental design in a previous study [5], the pairs are:

(1) [strange] / [same identity] / [nature] (A - A)
(2) [strange] / [same identity] / [nature and synthetic] (A - A')
(3) [strange] / [different identity] / [nature] (A - B)
(4) [familiar] / [same identity] / [nature] (C - C)
(5) [familiar] / [same identity] / [nature and synthetic] (C - C')
(6) [familiar] / [different identity] / [nature] (C - D)

A, B, C and D denote 4 different speakers: A and B are strange speakers to the participants, while C and D are familiar speakers. A' and C' represent synthetic voices of the same speakers as A and C, respectively.

Every stimulus can be categorized into one of three conditions, paired randomly with balanced frequency. Each pair will be heard for twice, and all the pairs are separated into 4 blocks with balanced sequence. The trial procedure is shown in Figure 3. Participants listen to a voice pair, waiting for the green fixation response with keyboard. After that, they are asked to choose from 1 to 7 to assess the degree of similarity.

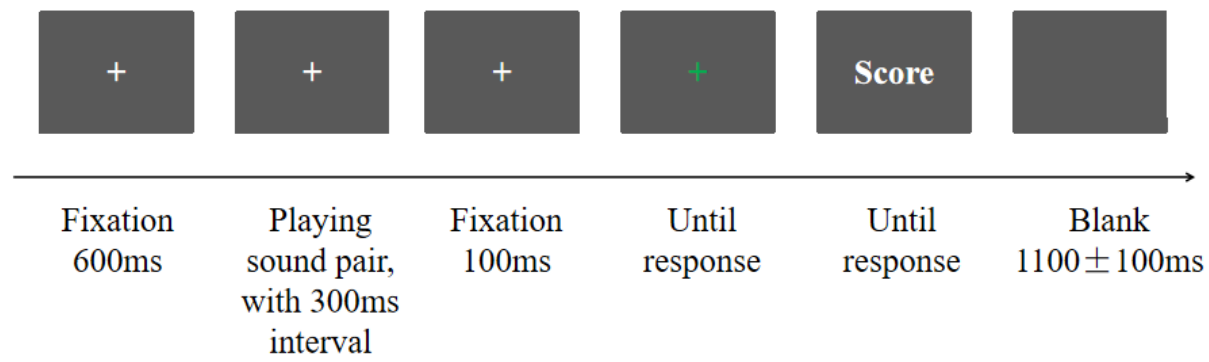

Figure 3: *Trial procedure of* speaker identity *similarity perception*

[1]https://github.com/Plachtaa/seed-vc

#### 2.3.2. *Intonation recognition*

For intonation recognition, the participants are asked to make a judgement on whether the intonation is statement or question. The 8 conditions are strange-synthetic-statement, strange-synthetic-question, strange-real-statement, strange-real-question, familiar-synthetic-statement, familiar-synthetic-question, familiar-real-statement, familiar-real-question. All of stimulus are separated in to 4 blocks. The procedure of trial is shown in Figure 4. Participants listen to a voice stimulus, waiting for the green fixation to make response through keyboard.

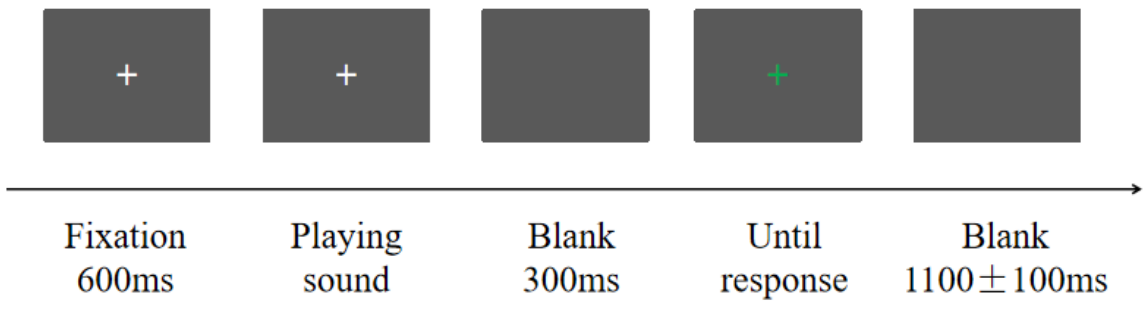

Figure 4: *Trial procedure of intonation recognition*

#### 2.3.3. *Experimental setting*

Participants listen to the voice stimuli in front of a screen monitor at 60 Hz with 1024*768 resolution. In both speaker identity similarity perceptions task and intonation recognition tasks, participants are asked to wear a headphone, then sitting 65cm from the computer, in a quiet laboratory environment. The voices stimuli are presented comfortably through the headphone. Participants are allowed to adjust the computer volume to a comfortable listening level. Figure 5 is the experimental setting.

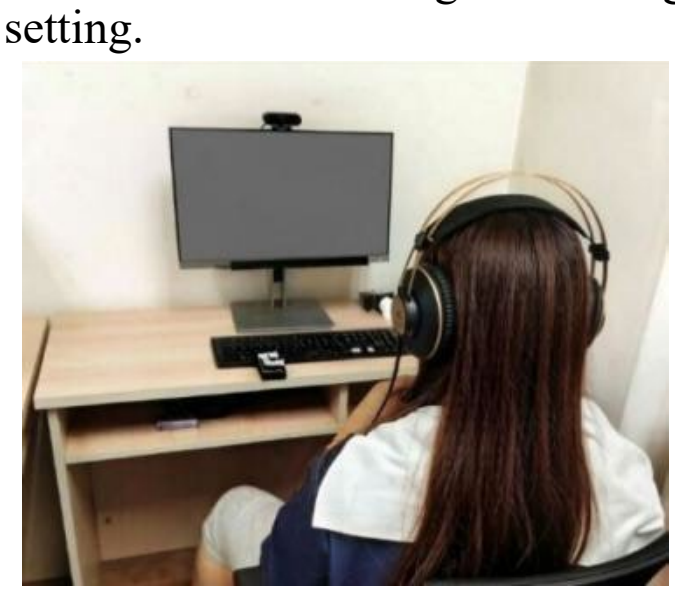

Figure 5: *The experimental setting*

### 2.4. Data analysis

Statistical analyses are conducted in R. For speaker identity similarity perception study, log-transformed RT data are analyzed with linear mixed-effects models implemented in the lme4 package [21]. ACC data are analyzed with generalized linear mixed-effects models with a binomial link function. Score data are analyzed using cumulative link mixed models (CLMMs) with a logit link function, implemented in the ordinal package in R. Given the ordinal nature of the 7-point Likert scale, CLMMs were used instead of linear models. Intonation, familiarity and speech type are entered as fixed effects, while participants and items are entered as random effects.

For intonation recognition study, RTs are log-transformed prior to analysis to reduce skewness and approximate normality. Log-transformed RTs were analyzed using linear mixed-effects models (LMMs) implemented in the lme4 package in R. Fixed effects included speech type, familiarity, and intonation, with random intercepts for participants and items. Accuracy data exists ceiling effect (mean accuracy is more than 95%). Due to potential separation issues in standard logistic regression models, accuracy data were analyzed using Firth's bias-reduced logistic regression implemented in the logistic package in R. The high accuracy suggest that the task was generally easy for adult participants, as expected. Therefore, reaction time measures provide a more sensitive index of processing differences across conditions.

## 3. Results

### 3.1. Speaker identity similarity perception

#### 3.1.1. *Accuracy*

Logistic regression revealed a significant effect of speech type (*Chisq*(1) = 11.244, $p < 0.001$), indicating higher accuracy for 2 natural speeches in one pair (Mean = 0.917, SE = 0.014) than natural and synthetic speech in one pair (Mean = 0.851, SE = 0.018; $z = 3.246$, $p = 0.001$). A significant interaction between speech type and intonation is found (*Chisq*(1) = 5.988, $p = 0.014$). Simple effects are then used to further explore it. For pairs consisted of only natural speeches, the accuracy of question stimuli (Mean = 0.946, SE = 0.014) is significantly higher than statement (Mean = 0.875, SE = 0.022; $z = 2.83$, $p = 0.046$). For pairs consisted of both natural and synthetic speeches, no significant difference exists between question (Mean = 0.826, SE = 0.025) and statement (Mean = 0.872, SE = 0.022; $z = -1.464$, $p = 0.143$). No other significant differences are observed.

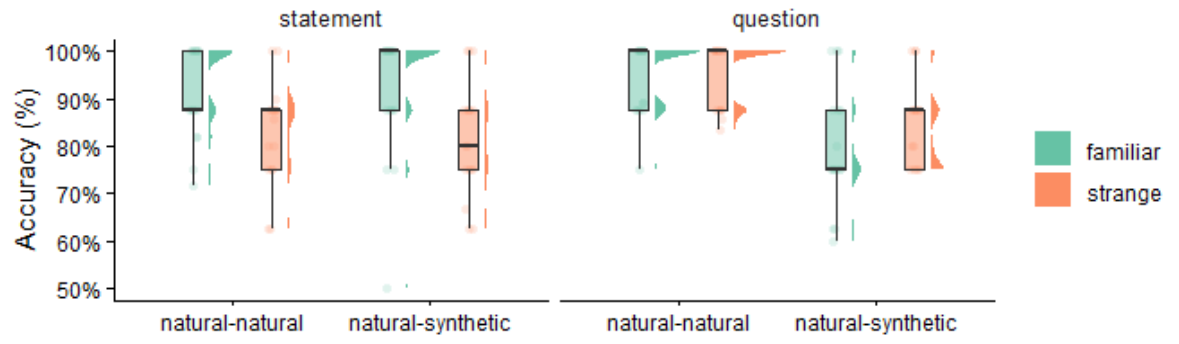

Figure 6: *Accuracy of speaker identity similarity perception*

#### 3.1.2. *Reaction time*

A significant effect of familiarity is observed ($F(1, 1052) = 6.713$, $p = 0.010$), indicating longer RTs for strange speakers (Mean = 0.848s, SE = 0.089) compared to familiar speakers (Mean = 0.748s, SE = 0.078). No other significant differences are observed.

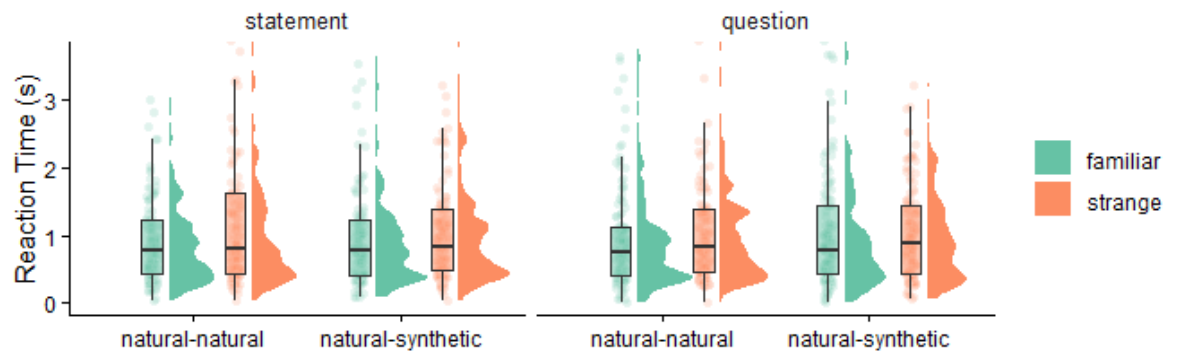

Figure 7: *Reaction time of speaker identity similarity perception*

#### 3.1.3. *Score of similarity*

A significant effect of familiarity is observed (*Chisq*(1) = 5.448, $p = 0.020$), indicating higher ratings for strange speakers (Mean = 3.930, SE = 0.131) than the familiar (Mean = 3.560, SE = 0.135; $z = -2.372$, $p = 0.018$). Another significant main effect is found in speech type (*Chisq*(1) = 3.906, $p = 0.048$), indicating that 2 natural speech types (Mean = 3.900, SE = 0.136) are scored higher in similarity than natural and synthetic speeches (Mean = 3.600, SE = 0.130; $z = 1.899$, $p = 0.058$). No other significant differences are observed.

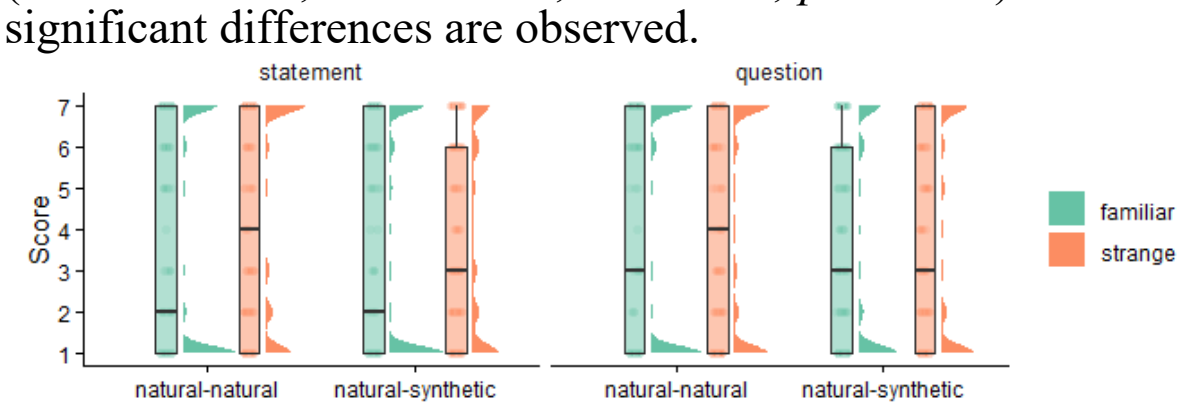

Figure 8: *Score of speaker identity similarity in similarity perception*

### 3.2. Intonation recognition

#### 3.2.1. *Accuracy*

Firth's bias-reduced logistic regression revealed a significant interaction of familiarity and speech type (*LR Chisq*(1) = 4.078, $p$ = 0.043). Simple effects are then used to further explore it. For natural speeches, the accuracy of strange voices (Mean = 0.987, SE = 0.006) is higher than the familiar (Mean = 0.982, SE = 0.008; $z$ = -0.468, $p$ = 0.640). For synthetic speeches, the accuracy of strange voices (Mean = 0.973, SE = 0.012) is lower than the familiar (Mean = 0.992, SE = 0.007; $z$ = 1.287, $p$ = 0.198). No other significant differences are observed.

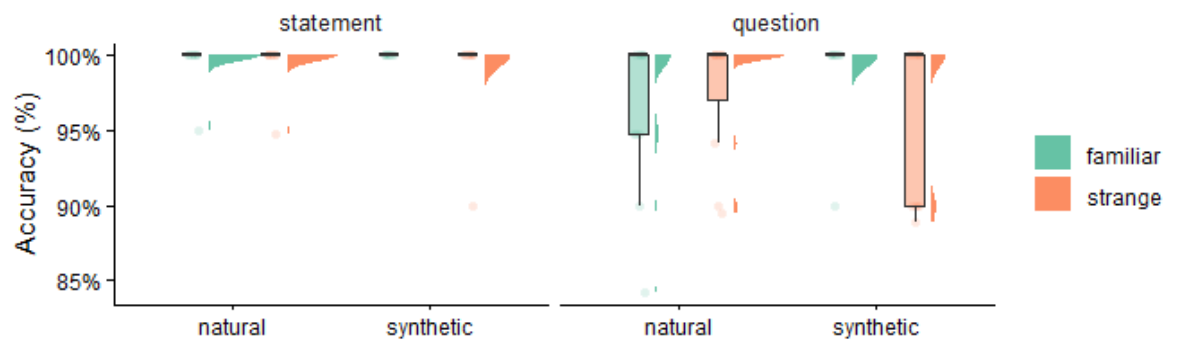

Figure 9: *Accuracy of intonation recognition*

#### *3.2.2 Reaction time*

A marginal significant effect of intonation is observed ($F$(1, 1178.03) = 3.844, $p$ = 0.050), indicating longer RTs for questions (Mean = 0.270s, SE = 0.014) compared to statement (Mean = 0.249s, SE = 0.013). No other significant differences are observed.

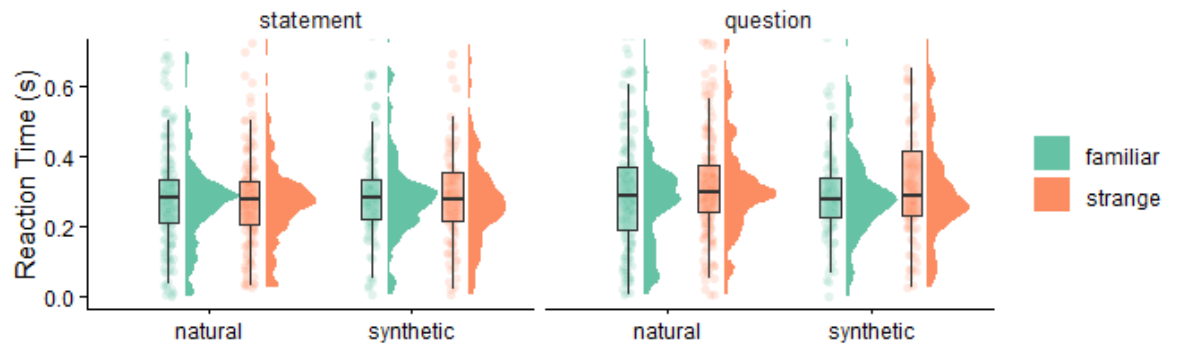

Figure 10: *Reaction time of intonation recognition*

## 4. Discussion

The present study aims to find out whether there is difference of voice processing between natural and synthetic voices, and how familiarity and intonation contribute to the process. In the speaker identity similarity perception task, an interaction towards accuracy between speech type and intonation is observed. In the intonation recognition task, an interaction towards accuracy between speech type and familiarity is observed. In the following sections, these findings will be interpreted respectively in different tasks.

### 4.1. Speaker identity similarity perception

Prior studies have noted the complexity of human speech, because of its tone, timbre, intonation, and tempo, as well as the potential quality compression when dealing with these details [12, 13]. An initial objective of the project is to identify whether there are significant differences between real voices and synthetic voices, which are strong enough to affect people's perception and recognition. The current study finds that there is a significant interaction between speech type and intonation. This suggested that question may serve as a cue for speaker identification, which in line with an opinion that linguistic prosody contributes to the impression of speakers identification [22, 23]. Meanwhile, may be further influenced by synthetic voices. The results are in accordance with recent studies indicating the difficulty in replicating those mentioned voices identifications [12, 13].

Another important finding is that familiarity shows significant main effect in reaction times and scoring. Participants make response to voices from familiar people faster than the strange, which supports the opinion that familiarity facilitates speech processing efficiency. Familiar voices benefit from stored acoustic representations in long-term memory, allowing listeners to process and recognize them more rapidly and with reduced cognitive effort [2]. Another result is that participants score the similarity of familiar people lower than the strange. A possible explanation for this might be that the ease of processing makes participants able to deal with acoustic representations from familiar people precisely. In other word, standard for evaluation differs. Faster reaction times indicate facilitated processing, while lower similarity ratings reflect more precise discrimination. This pattern is consistent with models of perceptual learning and talker adaptation, which propose that familiarity improves the resolution of perceptual representations and enhances sensitivity to within-speaker variation [24].

### 4.2. Intonation recognition

Several reports have shown that familiarity may conduct different contribution to voice processing in different situations [2, 24]. When recognizing the intonation, for accuracy, interaction between familiarity and speech type is found. Specifically, under the condition of voices from real people, the accuracy of the strange is higher than the familiar ones. Under the conditions of synthetic voices, the result is converse. This result may be explained by the different roles of familiarity. For the former result, to process natural voices, participants are able to consider more details of familiar voices, and making comparison between the present voices and the representations in their cognitive structure, that is, judging by different standards between natural voices and synthetic voices.

Meanwhile, in this study, for reaction time, the effect of intonation is marginal significant. This discrepancy could be attributed to the differences of F0, which seems to support the ideas that the difference of F0 between statement and question is more discernable at the end of sentences [25, 26, 27]. Since no significant interaction has been observed, it indicates that this phenomenon holds for both real and synthetic speech.

Based on the above findings, this study reveals that leveraging current AI voice cloning technology to construct high-quality corpora for language training remains challenging. It is particularly evident that there are limitations when the synthesis includes complex prosodic features (e.g. intonation) and social cues (e.g. speaker familiarity). However, this paper represents only a preliminary exploration of this question, whilst AI cloning technology continues to evolve. Future work may undertake more extensive and multidimensional investigations into this matter.

## 5. Conclusions

This study examines how speech type (natural and synthetic), familiarity (familiar and strange) and intonation (statement and question) influence participants' performance in similarity perception and intonation recognition tasks. In similarity perception task, significant interactions are found on intonation and speech type in accuracy, indicating that prosodic details, such as intonation may contribute to speakers identification. In intonation recognition task, significant interactions are found on familiarity and speech type in accuracy, indicating that participants may treat voices from familiar people more prudently. The results provide preliminary behavioral evidence for the perceptual viability of synthetic voices generated by SVC model in controlled experimental contexts. The present findings suggest that SVC-generated voices holds potential for corpus construction in intonation research, though further refinement is needed to fully capture speaker-level prosodic features. Future studies may extend these findings in a larger number of participants, collecting data from diverse participants with broader corpora.

## 6. Acknowledgements

This work was supported by the National Key R&D Program of China (2024YFC3308500), the NSFC (62332015), the project of China Disabled Persons Federation (CDPF2023KF00002), the Youth Innovation Promotion Association CAS Grant (2023119), and the China Postdoctoral Science Foundation (2024M763393). This study was approved by the Ethics Committee of the Institute of Software, Chinese Academy of Sciences (approval no. 2025-YF-001).

## 7. Generative AI Disclosure

Generative AI is only used for grammar checking and manuscript polishing. The authors have reviewed and edited the content, and are accountable for the work.